\pdfoutput=1
\documentclass[%
twocolumn,
showpacs,preprintnumbers,
 amsmath,amssymb,
 aps,
 prl,
floatfix,
]{revtex4-1}

\usepackage{graphicx}
\usepackage{dcolumn}
\usepackage{bm}
\usepackage[utf8]{inputenc}
\usepackage{amsmath}
\usepackage[amssymb]{SIunits}
\usepackage{amsfonts}
\usepackage{amssymb}
\usepackage{subfigure}
\usepackage{mathtools}
\usepackage{dsfont}
\usepackage{color}
\usepackage{isotope}
\usepackage{calc}
\usepackage{braket}

\begin{document}


\title{Temperature gradient driven lasing and stimulated cooling}

\author{K. Sandner}
\affiliation{Institute for Theoretical Physics, Universität Innsbruck, Technikerstrasse 25, 6020 Innsbruck, Austria}
\author{H. Ritsch}
\affiliation{Institute for Theoretical Physics, Universität Innsbruck, Technikerstrasse 25, 6020 Innsbruck, Austria}

\date{\today}

\begin{abstract}

A laser can be understood as thermodynamic engine converting heat to a coherent single
mode field close to Carnot efficiency. From this perspective spectral shaping of the
excitation light generates a higher effective temperature on the pump than on the gain
transition. Here, using a toy model of a quantum well structure with two suitably designed
tunnel-coupled wells kept at different temperature, we study a laser operated on an actual
spatial temperature gradient between pump and gain region.  We predict gain and narrow
band laser emission for a sufficient temperature gradient and resonator quality.  Lasing
appears concurrent with amplified heat flow and points to a new form of stimulated solid
state cooling. Such a mechanism could raise the operating temperature limit of quantum
cascade lasers by substituting phonon emission driven injection, which generates intrinsic
heat, by an extended model with phonon absorption steps.            


\end{abstract}

\maketitle


Lasing is facilitated by implementing gain within an optical resonator. For stimulated
emission based amplification this requires an inverted population distribution between the
amplifier quantum states~\cite{lamb1964theory}. In theoretical descriptions pumping to
generate inversion can be well described by coupling reservoirs of different temperature
to the pump, the injection and the lasing
transition~\cite{boukobza2006thermodynamic,youssef2010quantum}. Practically this requires
precise spectral design of the pump radiation to overlap mainly with the pump transition
with strong suppression on the lasing and injection transitions to yield inversion.
Interestingly, such inversion can be consistently mimicked by simply assuming an effective
negative temperature on the lasing transition~\cite{graham1970laserlight}. 
At optical frequencies almost no thermal photons are present at room temperature and
simple filtering of the pump light is sufficient. However, simple spectral shaping in the
far infrared or even THz regime is practically hampered by ubiquitous thermal quanta, which
create strong restrictions on the laser operating temperature~\cite{kumar20101}.

In this paper we study possibilities to directly use spatial temperature gradients instead
of spectral pump design to generate inversion and facilitate gain and lasing. Although we
mostly concentrate on the basic principles and mechanisms of such a setup, we also try to
connect to a possible real world implementation based on designed quantum well structures.
Here the optical properties and level structures in far infrared or even THz regime can be
precisely engineered. While such a temperature gradient laser at first seems more of
conceptual than of practical importance, the underlying principle can be easily
generalized to a system with different pump steps. In particular thermal or phonon
absorption based injection steps to the gain transition lead to interesting modification
of the thermal operation conditions of such a laser. Lasing can be tied to thermal
absorption and heat flow to avoid unwanted heating and control heat
flow~\citep{epstein1995observation}.

\begin{figure}[t]
	\centering
	\includegraphics[width=0.4\textwidth]{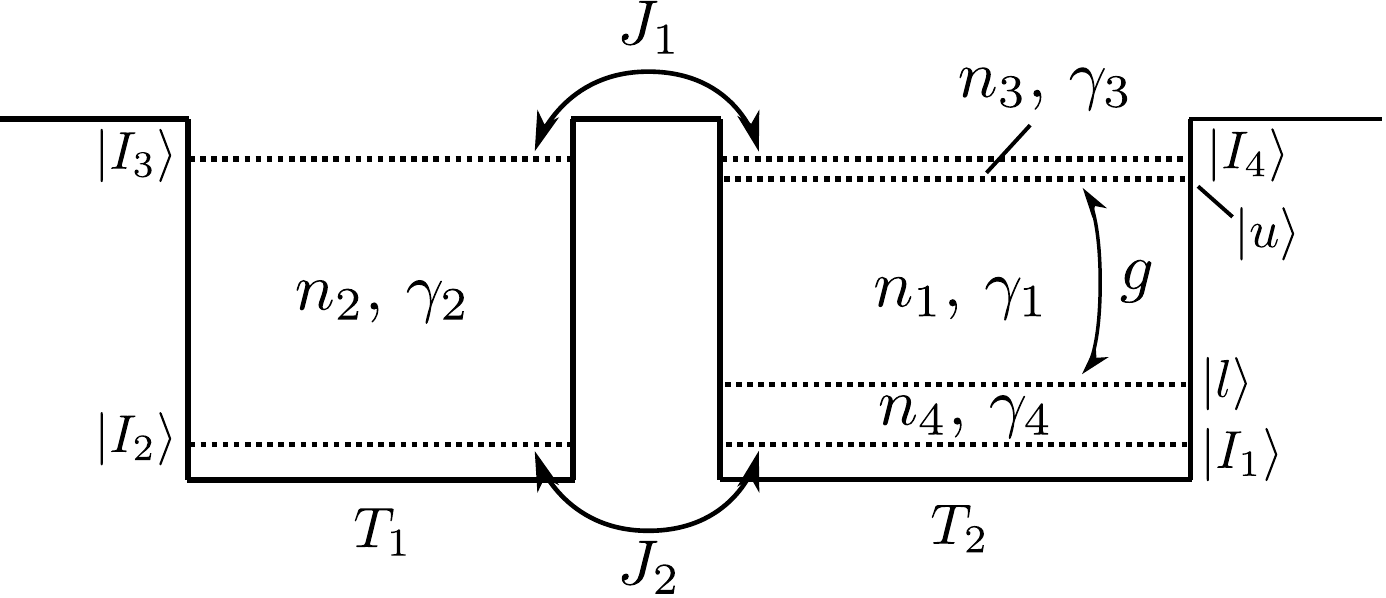}
	\caption{Sketch of the two-well structure and the energy levels. The wave functions are
	assumed to be well localized in each well while the coupling is described by the
	tunneling constants $J_{1}$ and $J_{2}$.}
	\label{fig:wells}
\end{figure}

The model we study is a simple structure comprised of two quantum wells coupled to a
cavity. A sketch is shown in Fig.~\ref{fig:wells}.
The Hamiltonian describing a system of $N$ two-well structures coupled to a single cavity
mode is given by $\operatorname{H}=\operatorname{H}_{0}+\operatorname{H}_{\text{I}}$
with
\begin{multline}
	\operatorname{H}_{0}=\hbar \omega_{m} a^{\dagger}a+\hbar
	\sum^{N}_{j=1}\Bigl[\omega_{u}\ket{u_{j}}\bra{u_{j}}+
	\omega_{l}\ket{l_{j}}\bra{l_{j}}\\
	+\sum^{4}_{k=1} \omega_{I_{k}}\ket{I_{k,j}}\bra{I_{k,j}}\Bigr]
	\label{eq:hamiltonian0}
\end{multline}
and the interaction Hamiltonian
\begin{align}
	\operatorname{H}_{\text{I}}&=\hbar g \sum^{N}_{j=1}\left( a  \ket{u_{j}}\bra{l_{j}}+a^{\dagger}
	\ket{l_{j}}\bra{u_{j}}\right)\notag\\
	&+\hbar\sum^{N}_{j=1}\left( J_{1}
	\ket{I_{4,j}}\bra{I_{3,j}}+ J_{2}
	\ket{I_{2,j}}\bra{I_{1,j}}+h.c.\right)\ .
	\label{eq:hamiltonianI}
\end{align}
The energy eigenstates are denoted by $\ket{I_{1,j}},\dots, \ket{I_{4,j}}$, $\ket{u_{j}}$ and
$\ket{l_{j}}$ with the eigenenergies $\hbar \omega_{I_{1}},\dots,\hbar \omega_{I_{4}}$, $\hbar
\omega_{u}$ and $\hbar \omega_{l}$ and $a^{\dagger}$ is the creation operator for a photon
in the cavity mode with frequency $\omega_{m}$. The coupling between the $u \rightarrow l$ transition
and the mode is described by the coupling rate $g$. Tunneling between the states
$\ket{I_{1,j}}$ and $\ket{I_{2,j}}$ and the states $\ket{I_{3,j}}$
and $\ket{I_{4,j}}$ is controlled by the tunneling matrix elements $J_{1}$ and $J_{2}$. For
the forthcoming analysis we define all frequencies with respect to $\omega_{l}$ and switch
to a frame rotating with $\omega_{m}$ introducing the detuning $\Delta_{u}=\omega_{u}-\omega_{m}$.
The dynamics of the reduced density matrix is described by the master equation
\begin{align}
	\frac{\mathrm{d}}{\mathrm{d}t}\rho=\frac{\mathrm{i}}{\hbar}\left[
\rho,\operatorname{H} \right]+\mathcal{L}\left[ \rho \right]
	\label{eq:master}
\end{align}
where the Liouvillian $\mathcal{L}$ describes dissipative processes, including the
coupling to the environment at finite temperature, see supplementary material. 

The decay rate of the cavity is denoted by $\kappa$, while the spontaneous emission rates
of the remaining transitions are denoted by $\gamma_{1},\dots,\gamma_{4}$. The thermal
photon number at the transition frequencies $\omega_{1}=\omega_{u}-\omega_{l}$,
$\omega_{2}=\omega_{I_{3}}-\omega_{I_{2}}$, $\omega_{3}=\omega_{I_{4}}-\omega_{u}$ and
$\omega_{4}=\omega_{l}-\omega_{I_{1}}$ at temperature $T$ is denoted by
$n_{i}\left(T\right)$ ($i=1,\dots, 4$).

We assume that we have a temperature gradient across a sample containing $N$ two-well
structures and that $T_{1}>T_{2}$ in each structure. As $T_{1}\neq 0$ we will find
population in the state $\ket{I_{3}}$. Via the tunnel-coupling $J_{1}$ the state
$\ket{I_{2}}$ gets populated as well. Due to the lower temperature $T_{2}$ of the second
well the population is preferably transferred to the state $\ket{u}$. With a certain
probability a photon is created in the cavity and leaves it. The cycle ends with the
tunneling from state $\ket{I_{1}}$ to $\ket{I_{2}}$.


Using Eq.~\eqref{eq:master} we calculate operator expectation value equations for the
populations in each state and the quantities they couple to \citep{carmichael1993open}.
The equations are truncated to include only second order correlation functions. The
population of the state $\ket{u_{j}}$ is described by
\begin{align}
	\frac{\mathrm{d}}{\mathrm{d}t}\left< \ket{u_{j}}\bra{u_{j}} \right>&=
	-\mathrm{i}g\left(\left< a\ket{u_{j}}\bra{l_{j}} \right>-\left< a^{\dagger}\ket{l_{j}}\bra{u_{j}} \right>
	 \right)\notag\\
	 &-\left( 2\gamma_{1}\left( n_{1}+1 \right)+2\gamma_{3}n_{3} \right)\left<
	 \ket{u_{j}}\bra{u_{j}} \right>\notag\\
	 &+2 \gamma_{1}n_{1}\left< \ket{l_{j}}\bra{l_{j}} \right>\notag\\
	 &+2 \gamma_{3}\left( n_{3}+1 \right)\left< \ket{I_{4,j}}\bra{I_{4,j}} \right>\ ,
	\label{eq:u_j}
\end{align}

where we omitted the temperature dependence of the $n_{i}$ and used the assumption of
total phase invariance ($\left< a \right>=\left< \ket{u_{j}}\bra{l_{j}} \right>=0$) to
simplify the expansion of third order expectation values~\cite{kubo1962generalized}. This leads to the next
equation
\begin{multline}
	\frac{\mathrm{d}}{\mathrm{d}t}\left< a \ket{u_{j}}\bra{l_{j}} \right>=	\mathrm{i}\Delta_{u}\left< a \ket{u_{j}}\bra{l_{j}} \right>\\
	\shoveleft-\mathrm{i}g\left( \left[1+\left<  a^{\dagger}a\right>\right]\left< \ket{u_{j}}\bra{u_{j}}\right>-\left<
	a^{\dagger}a\right> \left< \ket{l_{j}}\bra{l_{j}}\right> \right)\\
	-\left(\kappa+\gamma_{1}\left( 2n_{1}+1 \right)+\gamma_{3}n_{3}+\gamma_{4}\left(
	n_{4}+1
	\right)  \right)\left< a \ket{u_{j}}\bra{l_{j}} \right>\\
	-\mathrm{i}g\left( N-1 \right)\left<  \ket{l_{j}}\bra{u_{j}}\ket{u_{k}}\bra{l_{k}}\right>
	\label{eq:au_jl_j}
\end{multline}
which describes the rate at which excitations are transferred between the cavity and the
structure. The number of photons follows
	$\frac{\mathrm{d}}{\mathrm{d}t}\left< a^{\dagger}a \right>=-2\kappa\left< a^{\dagger}a
	\right> +2 \kappa n_{1}\notag-\mathrm{i}gN\left(  \left< a^{\dagger}
	\ket{l_{j}}\bra{u_{j}} \right>-\left< a \ket{u_{j}}\bra{l_{j}} \right>\right)$.
In the last line of Eq.~\eqref{eq:au_jl_j} we see that cooperative effects between the
structures emerge as a result of the common interaction with the cavity. We
find
\begin{multline}
	\frac{\mathrm{d}}{\mathrm{d}t}\left<
	\ket{l_{j}}\bra{u_{j}}\ket{u_{k}}\bra{l_{k}}\right>=-\mathrm{i}g\left[ \left( \left<
	a^{\dagger}
	\ket{l_{j}}\bra{u_{j}} \right>-\left< a \ket{u_{j}}\bra{l_{j}} \right>
	\right)\right.\\
	\left. \times \left(  \left< \ket{u_{j}}\bra{u_{j}} \right> -\left<
	\ket{l_{j}}\bra{l_{j}} \right>\right)\right]\\
	-\left( 2 \gamma_{1}\left( 2 n_{1}+1 \right)+2 \gamma_{3}n_{3}+2 \gamma_{4}\left(
	n_{4}+1
	\right) \right)\\
	\times\left<
	\ket{l_{j}}\bra{u_{j}}\ket{u_{k}}\bra{l_{k}}\right>\ .
	\label{eq:l_ju_ju_kl_k}
\end{multline}
In addition we calculate equations for $\left< \ket{l_{j}}\bra{l_{j}}\right>$, $\left<
\ket{I_{i,j}}\bra{I_{i,j}}\right>$ ($i=1,2,3,4$), $\left<
\ket{I_{1,j}}\bra{I_{2,j}}\right>$ and $\left< \ket{I_{3,j}}\bra{I_{4,j}}\right>$.

\begin{figure}[t]
	\centering
	\includegraphics[width=0.5\textwidth]{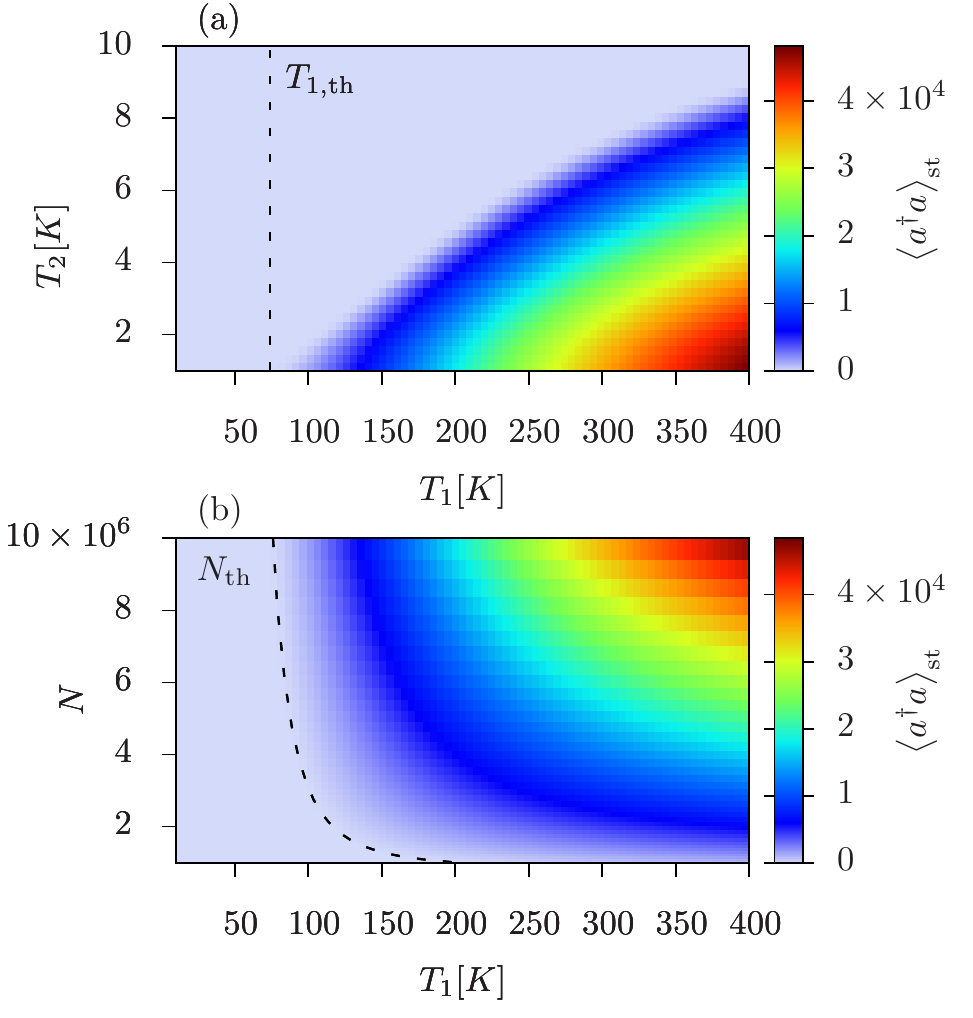}
	\caption{(Color online) (a) Numerical solution for the steady state number of photons
	$\left< a^{\dagger}a \right>_{\text{st}}$ in the cavity as a function of $T_{1}$ and
	$T_{2}$. The number of structures was chosen to $N=10^{7}$. The dashed line indicates
	$T_{1,\text{th}}$ necessary to obtain cooperative behavior (valid for $T_{2}$ small
  enough to justify setting $n_{1}=n_{3}=n_{4}=0$). (b) $\left< a^{\dagger}a
	\right>_{\text{st}}$ as a function of $T_{1}$ and $N$ ($T_{2}=\unit{0.1}{\kelvin}$). The
	analytical solution gives $N_{\text{th}}\left( T_{1} \right)$ (dashed line) and reliably
	predicts the onset of cooperative behavior and hence a significant increase in $\left<
	a^{\dagger}a \right>_{\text{st}}$.}
	\label{fig:laser_photons}
\end{figure}

Approximate analytical solutions can be deduced under certain assumptions and give insight
into the main mechanisms~\citep{meiser2009prospects}. We assume that $T_{2}$ is small
enough to justify setting $n_{1}=n_{3}=n_{4}=0$ and that $\Delta_{u}=0$. By keeping only
the decay term proportional to $\kappa$ and the last line in Eq.~\eqref{eq:au_jl_j}, where
we approximate $N-1\approx N$, we obtain the steady-state
\begin{align}
	\left< a \ket{u_{j}}\bra{l_{j}} \right>_{\text{st}}= -\frac{\mathrm{i}g N}{\kappa}\left<
	\ket{l_{j}}\bra{u_{j}}\ket{u_{k}}\bra{l_{k}}\right>_{\text{st}}\ .
	\label{eq:approx_au_jl_j_st}
\end{align}
Under the assumption of negligible $T_{2}$ and by using Eq.~\eqref{eq:approx_au_jl_j_st} it
is possible to calculate approximate steady state solutions for all of the involved
quantities. The  steady state of the correlation between the structures
$\left<\ket{l_{j}}\bra{u_{j}}\ket{u_{k}}\bra{l_{k}}\right>_{\text{st}}$ is of particular
interest. It shows for which parameters the formation of cooperative behavior with the
mode is possible~\citep{meiser2009prospects}.
For a given number of structures $N$ we can calculate the critical temperature
$T_{1,\text{th}}$ that is necessary to meet the condition for cooperative behavior
($\left<\ket{l_{j}}\bra{u_{j}}\ket{u_{k}}\bra{l_{k}}\right>_{\text{st}}>0$). At the same
time, for a given temperature $T_{1}$ and the resulting number $n_{2}\left( T_{1} \right)$
a certain threshold number $N_{\text{th}}$ of structures is necessary for cooperative
behavior. Details can be found in the supplementary material.

The full set of equations is also integrated numerically to determine the exact steady
states and to compare them to the approximate solutions. The parameters $\kappa=2 \pi
\times 5 \times \unit{10^{5}}{\hertz}$ (i.e. FWHM linewidth of $\unit{1}{\mega\hertz}$), $g=(\kappa/3)
\times 10^{-3}$, $\Delta_{u}=0$ ,$\gamma_{1}=2 \pi \times 4 \times \unit{10^{2}}{\hertz}$,
$\gamma_{2}=\gamma_{3}=\gamma_{4}=\gamma_{1} \times 10^{2}$,
$\omega_{1}= 2 \pi\times \unit{7.9}{\tera\hertz}$, $\omega_{2}=2 \pi \times \unit{9}{\tera\hertz}$,
$\omega_{3}=2 \pi \times \unit{0.1}{\tera\hertz}$, $\omega_{4}= 2 \pi \times \unit{1}{\tera\hertz}$   remain fixed for the
forthcoming calculations, while $N$, $T_{1}$ and $T_{2}$ are varied. The number of photons
as a function of $T_{1}$ and $T_{2}$ for $N=10^{7}$ is shown in
Fig.~\ref{fig:laser_photons}(a). The occupation of the cavity mode strongly depends on
the temperature gradient between the wells. The approximative analytical treatment for
small $T_{2}$ gives the threshold temperature $T_{1,\text{th}}$ that signals the onset of
cooperative behavior. For $T_{2}> \unit{1}{\kelvin}$ the approximative solutions break
down. In Fig.~\ref{fig:laser_photons}(b) we show $\left< a^{\dagger}a \right>_{\text{st}}$ as a
function of $T_{1}$ and $N$ for fixed $T_{2}=\unit{0.1}{\kelvin}$.  Here, the approximative
analytical solutions give a reliable result for which parameters to expect a high number
of photons in the cavity. The numerical solutions show that the number of photons in the
cavity increases rapidly as soon as we exceed $N_{\text{th}}$ for a given value of
$T_{1}$.

\begin{figure}[tbp]
	\centering
	\includegraphics[width=0.4\textwidth]{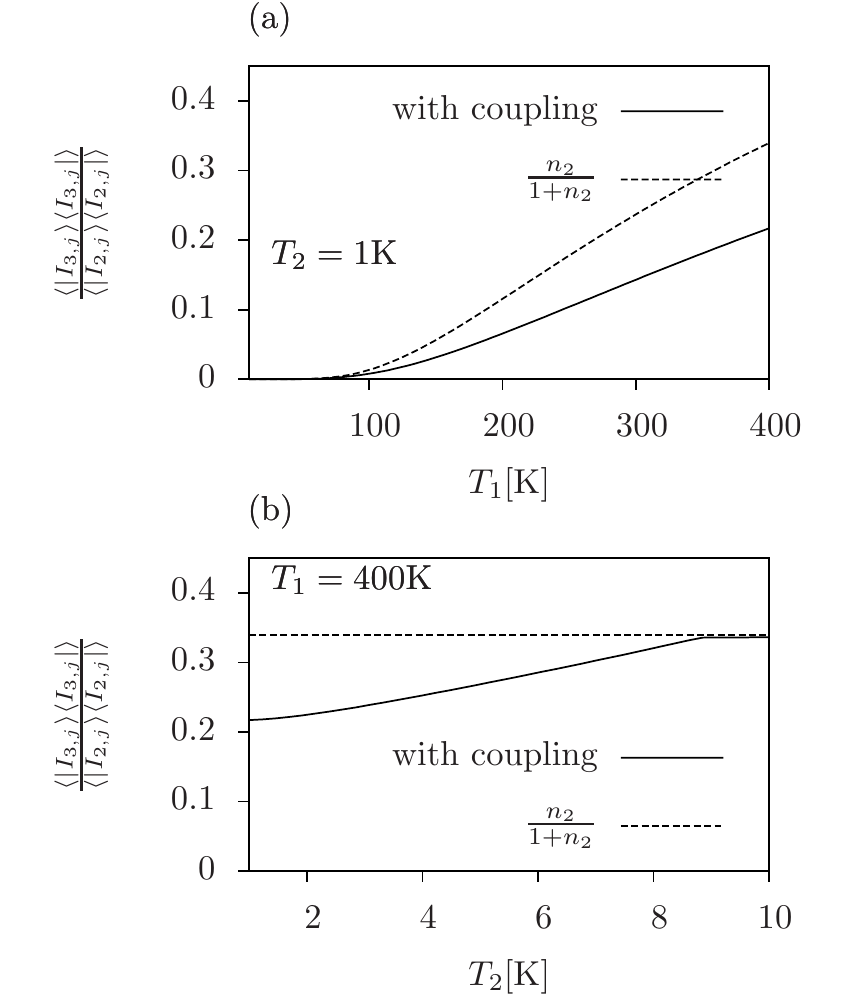}
	\caption{Occupation ratio in the first quantum well for the coupled and uncoupled case versus
	$T_{1}$ and $T_{2}$. In (a) we choose $T_{2}=\unit{1}{\kelvin}$ and vary $T_{1}$. The
	ratio between the population of $I_{3,j}$ and $I_{2,j}$ (solid line) increases with
	$T_{1}$ but remains below the dashed line which gives the ratio in the uncoupled case.
	In (b) the temperature of the hot part $T_{1}$ is fixed to $\unit{400}{\kelvin}$. For
	$T_{2}< \unit{9}{\kelvin}$ we again see the reduction of $\left<
	\ket{I_{3,j}}\bra{I_{3,j}} \right>/\left< \ket{I_{2,j}}\bra{I_{2,j}} \right>$.}
	\label{fig:cooling_first_well}
\end{figure}

Assume that we have a layer containing $N$ structures which initially is hot. If one side
of the layer is cooled by some means, the resulting temperature gradient will lead to a
population of the cavity mode.  The rate at which photons are lost from the cavity is
$2 \kappa \left< a^{\dagger}a \right>$.  This suggests that the cavity can serve as a
channel for cooling.  Without the second quantum well, or equivalently with
$J_{1}=J_{2}=0$, we find the ratio between the upper level $I_{3}$ and the lower energy
level $I_{2}$ in the first well to be $n_{2}/(1+n_{2})$.
In the presence of the second well, and in the regime where we find a considerable number
of photons in the cavity, this ratio is reduced.
Figure~\ref{fig:cooling_first_well}(a) shows how the ratio $\left<
\ket{I_{3,j}}\bra{I_{3,j}} \right>/\left< \ket{I_{2,j}}\bra{I_{2,j}} \right>$ changes with
increasing $T_{1}$ while $T_{2}=\unit{1}{\kelvin}$ remains fixed.
With increasing $T_{1}$ the relative occupation of the upper level $\ket{I_{3,j}}$
increases but remains below the result for the uncoupled case. The difference starts to
show at $T_{1}\approx \unit{75}{\kelvin}$ which is the critical temperature above which we
obtain a significant increase in $\left< a^{\dagger}a \right>_{\text{st}}$, as can be
seen in Fig.~\ref{fig:laser_photons}(a). The same comparison can be made for fixed
$T_{1}=\unit{400}{\kelvin}$ (leaving $n_{2}/(1+n_{2})$ fixed) and varying $T_{2}$. While
$T_{2}$ remains low enough to ensure a significant number of photons in the cavity, and
hence a considerable dissipation via the cavity mirrors, the occupation ratio in the first
well is reduced. This corresponds to an effectively lower temperature of the first well.

\begin{figure}[t]
	\centering
	\includegraphics[width=0.5\textwidth]{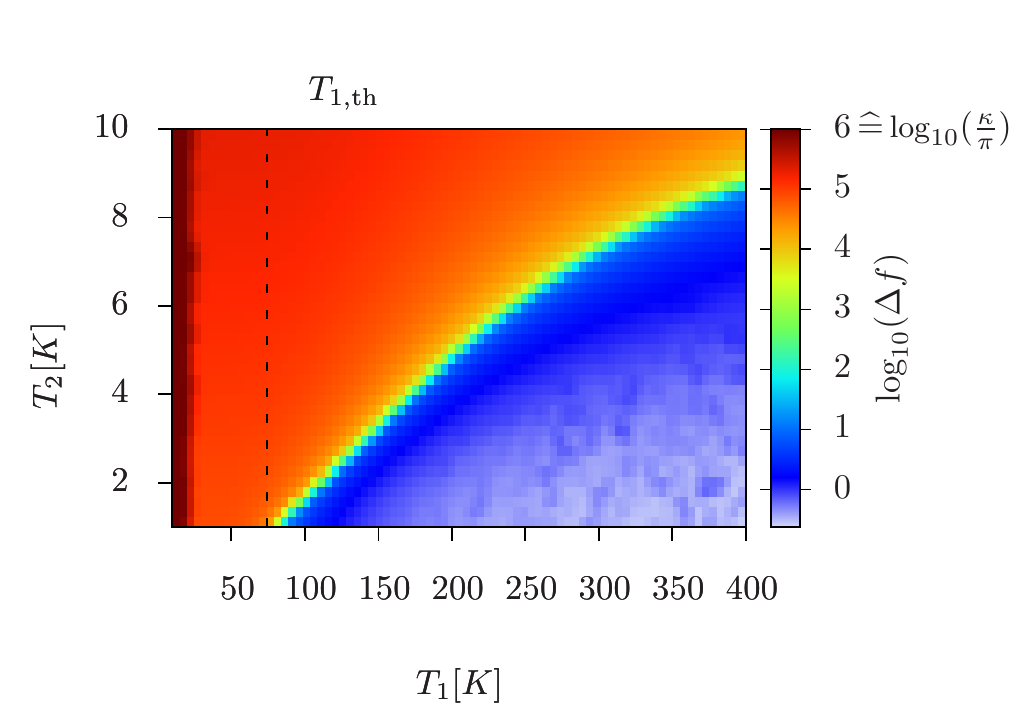}
	\caption{(Color online) Linewidth of the cavity output spectrum for varying $T_{1}$ and
	$T_{2}$ with $N=10^{7}$ (plotted on a logarithmic scale). The linewidth (FWHM) of the
	empty cavity $2 \frac{\kappa}{2 \pi}=\unit{1}{\mega\hertz}$ corresponds to the dark red
	color (upper boundary of the scale). The linewidth $\Delta f$ falls slightly below that limit with increasing $T_{1}$
	until it finally drops off dramatically as the cooperative effects emerge. For small
	$T_{2}$ the onset of cooperative behavior is marked by
	$T_{1,\text{th}}$.}
	\label{fig:lw_T1_T2}
\end{figure}

We can obtain a large number of photons in the cavity for $N>N_{\text{th}}\left( T_{1}
\right)$ and an appropriate ratio between $T_{1}$ and $T_{2}$. The temperature gradient
can therefore be seen as an effective pump. To characterize the emission we calculate the
spectrum of the cavity output field and the linewidth.  Using the quantum regression
theorem we obtain the differential equation for the first order correlation function
$\left< a^{\dagger}\left( t+\tau \right)a\left( t \right) \right>$ and $\left<
\ket{u_{j}}\bra{l_{j}}\left( t+\tau \right) a\left( t \right)\right>$, see supplementary
material.  To find the spectrum we calculate the Laplace transform of the coupled
equations ~\citep{meystre2007elements}. The steady state values of the involved quantities
are obtained numerically. In Fig.~\ref{fig:lw_T1_T2} the linewidth (FWHM) of the cavity
output spectrum is depicted. With the onset of cooperative behavior the linewidth is
reduced dramatically. Similar behavior is found for fixed $T_{2}$ and varying $T_{1}$ and
$N$.

The two-well system discussed so far clearly exhibits the essential physical mechanism of
heat to coherent light conversion in its purest form.  In the following we show that the
principle can be extended to multi step excitation processes, where several heat absorbing
steps are combined to lead to operation at higher emission frequencies. Mathematically,
this can be easily facilitated by incorporating an absorptive step in the second well by
choosing $\omega^{\prime}_{u}>\omega_{4}$, without needing a more complex model.  In the
dissipative dynamics we simply have to exchange the rates $\gamma_{3}n_{3}\left( T_{2}
\right)\leftrightarrow \gamma_{3}\left( n_{3}\left( T_{2} \right) +1 \right)$. Physically,
this means that in each cycle in this system the absorption of an extra thermal photon or
phonon with frequency $\omega^{\prime}_{u}-\omega_{I_{4}}$ is necessary. This implies that
we find a minimal temperature below which not enough phonons are present and lasing is
impossible. In general one could of course even think of a combination of thermally
assisted and externally driven excitation steps. We will refrain from this here to keep
the complexity of the model limited.

\begin{figure}[tbp]
	\centering
	\includegraphics[width=0.5\textwidth]{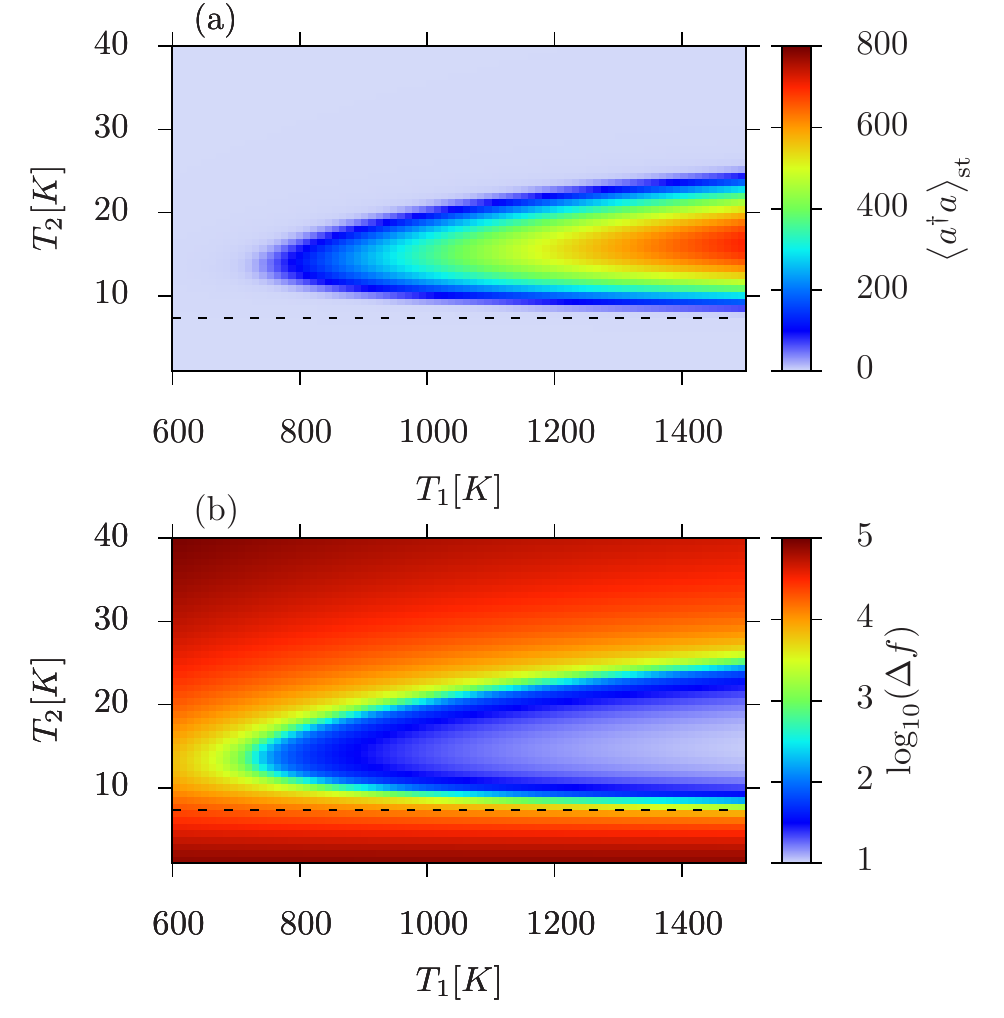}
	\caption{(Color online) (a) Number of photons in the mode versus $T_{1}$ and
	$T_{2}$.  For $T_{2}>T_{2,\text{th}}$ we find a sudden increase in
	$\left< a^{\dagger}a \right>_{\text{st}}$, and in (b) a corresponding drop in the linewidth of
	the cavity emission. The parameters remain unchanged except for
	$\gamma_{3}=\gamma_{1}=2 \pi \times 4 \times 10^{2}$ and
	$\omega^{\prime}_{u}=\omega_{u}+2 \pi \times \unit{0.2}{\tera \hertz}$.}
	\label{fig:switch_U_I4_photons_linewidth}
\end{figure}

Let us look at the results now. The temperature range for which $\left< a^{\dagger}a
\right>_{\text{st}}$ becomes significant is depicted in
Fig.~\ref{fig:switch_U_I4_photons_linewidth}(a). The dashed line marks the critical
temperature necessary for cooperative behavior, which is obtained analytically with
similar approximations as described above. The linewidth of the emission is shown in
Fig.~\ref{fig:switch_U_I4_photons_linewidth}(b). This concept could be incorporated as an
extra step in conventional quantum well lasers, which generate intrinsic heat during
operation, to extend their range of operation as the absorption of phonons would
counteract the heating process.


Our theoretical calculations exhibit the possibility to use a spatial temperature gradient
in a suitable active medium to induce optical gain and lasing.  While not very practicable
in the visible regime, a semiconductor heterostructure can be envisaged to generate
narrow bandwidth emission through a resonant cavity mode in the THz regime, where the
coherent emission of photons is sustained by the spatial  temperature gradient.  The
effective extraction of thermal energy from the system in parallel provides for stimulated
cooling of the first, hot quantum well. To use a similar idea at higher frequencies
without too high temperatures, a multi step excitation process can be envisaged, where some
steps could also invoke optical pumping or current injection. In general, introducing an
absorptive step  well before carrier injection to the upper laser level relies on
absorption of thermal quanta and therefore counteracts heating during operation. This
leads to a higher operating temperature limit. While we here only worked with an
oversimplified generic model, the physical principles behind should stay valid for more
realistic and thus more complex descriptions. In addition, transverse spatial temperature
gradients or even spatially separated structures connected by conducting wires might
provide for alternative configurations at greater length scales.\par

K.S. was supported by the DOC-fFORTE doctoral program of the ÖAW. H.R. acknowledges
support from the Austrian Science Fund FWF grant S4013.

%

\section*{Supplementary material}
\subsection*{Liouvillian}
The Liouvillian that describes dissipative processes, including the
coupling of the cavity and the structures to the environment at finite temperature, reads
\begin{widetext}
\begin{multline}
	\mathcal{L}\left[ \rho \right]=\kappa \left( n_{1}\left(T_{2}\right)+1 \right)\left( 2 a \rho a^{\dagger }-a^{\dagger }a
	\rho -\rho a^{\dagger }a \right)
	+\kappa n_{1}\left(T_{2}\right)\left( 2 a^{\dagger }\rho a -a a^{\dagger } \rho- \rho a a^{\dagger
	}\right)\\
	\shoveleft{+\gamma_{1}\left( n_{1}\left(T_{2}\right) +1 \right)\sum_{j}\left( 2 \ket{l_{j}}\bra{u_{j}} \rho
	\ket{u_{j}}\bra{l_{j}}\right.}
	\left. - \ket{u_{j}}\bra{u_{j}}\rho - \rho \ket{u_{j}}\bra{u_{j}}\right)\\
		\shoveleft{+\gamma_{1}\left( n_{1}\left(T_{2}\right) \right)\sum_{j}\left( 2 \ket{u_{j}}\bra{l_{j}} \rho
		\ket{l_{j}}\bra{u_{j}}\right.}
	\left. - \ket{l_{j}}\bra{l_{j}}\rho - \rho \ket{l_{j}}\bra{l_{j}}\right)\\
		\shoveleft{+\gamma_{2}\left( n_{2}\left(T_{1}\right) +1 \right)\sum_{j}\left( 2 \ket{I_{2,j}}\bra{I_{3,j}} \rho
		\ket{I_{3,j}}\bra{I_{2,j}}\right.}
	\left. - \ket{I_{3,j}}\bra{I_{3,j}}\rho - \rho \ket{I_{3,j}}\bra{I_{3,j}}\right)\\
		\shoveleft{+\gamma_{2}\left( n_{2}\left(T_{1}\right) \right)\sum_{j}\left( 2 \ket{I_{3,j}}\bra{I_{2,j}} \rho
		\ket{I_{2,j}}\bra{I_{3,j}}\right.} 
	\left. - \ket{I_{2,j}}\bra{I_{2,j}}\rho - \rho \ket{I_{2,j}}\bra{I_{2,j}}\right)\\
 	\shoveleft{+\gamma_{3}\left( n_{3}\left(T_{2}\right) +1 \right)\sum_{j}\left( 2 \ket{u_{j}}\bra{I_{4,j}} \rho
	\ket{I_{4,j}}\bra{u_{j}}\right.} 
	\left. - \ket{I_{4,j}}\bra{I_{4,j}}\rho - \rho \ket{I_{4,j}}\bra{I_{4,j}}\right)\\
		\shoveleft{+\gamma_{3}\left( n_{3}\left(T_{2}\right)\left(T_{2}\right) \right)\sum_{j}\left( 2 \ket{I_{4,j}}\bra{u_{j}} \rho
		\ket{u_{j}}\bra{I_{4,j}}\right.}
	\left. - \ket{u_{j}}\bra{u_{j}}\rho - \rho \ket{u_{j}}\bra{u_{j}}\right)\\
 	\shoveleft{+\gamma_{4}\left( n_{4}\left(T_{2}\right) +1 \right)\sum_{j}\left( 2 \ket{I_{1,j}}\bra{l_{j}} \rho
	\ket{l_{j}}\bra{I_{1,j}}\right.} 
	\left. - \ket{l_{j}}\bra{l_{j}}\rho - \rho \ket{l_{j}}\bra{l_{j}}\right)\\
		\shoveleft{+\gamma_{4}\left( n_{4}\left(T_{2}\right) \right)\sum_{j}\left( 2 \ket{l_{j}}\bra{I_{1,j}} \rho
		\ket{I_{1,j}}\bra{l_{j}}\right.} 
	\left. - \ket{I_{1,j}}\bra{I_{1,j}}\rho - \rho \ket{I_{1,j}}\bra{I_{1,j}}\right)\ .
\end{multline}\label{eq:dissip}
\end{widetext}

\subsection*{Approximate Analytical Solutions}
For $T_{2}$ small enough we can neglect $n_{1}$, $n_{3}$ and $n_{4}$. Inserting the
steadystate solutions for $\left< a \ket{u_{j}}\bra{l_{j}} \right>_{\text{st}}$, $\left<
\ket{u_{j}}\bra{u_{j}} \right>_{\text{st}}$ and $\left< \ket{l_{j}}\bra{l_{j}} \right>_{\text{st}}$
into Eq.~\eqref{eq:l_ju_ju_kl_k} we obtain
\begin{widetext}
\begin{multline}
	2 \left<\ket{l_{j}}\bra{u_{j}}\ket{u_{k}}\bra{l_{k}}\right>_{\text{st}}
	\Bigg[\gamma_{1} + \gamma_4\\
	+ \bigg( g^{4} N^{2} \left| J_{2}\right|^{2} 
	\left<\ket{l_{j}}\bra{u_{j}}\ket{u_{k}}\bra{l_{k}}\right>_{\text{st}} (2 +   3 n_2)
	\gamma_2 \gamma_3 (\gamma_2 +   n_2 \gamma_2 + \gamma_3) \gamma_4  + \left|
	J_{1}\right|^{2} \Big(g^{4} N^{2}
	\left<\ket{l_{j}}\bra{u_{j}}\ket{u_{k}}\bra{l_{k}}\right>_{\text{st}}
	n_2^2  \gamma_2^2 \gamma_3 \gamma_4 \\
	+\left| J_{2}\right|^{2} \big(2 g^{4} N^{2}
	\left<\ket{l_{j}}\bra{u_{j}}\ket{u_{k}}\bra{l_{k}}\right>_{\text{st}} ((\gamma_2 +
	\gamma_3) \gamma_4 + n_2 \gamma_2 (\gamma_3 + 2 \gamma_4)) +   g^2 N n_2  \gamma_2
	\gamma_3 (\gamma_1 - \gamma_4) \kappa +  n_2 \gamma_2 \gamma_3 (\gamma_{4}^2 -
	\gamma_{1}^2)
	\kappa^2\big)\Big)\bigg)\\
	/\bigg(\gamma_1 \kappa^2 \Big(\left| J_{2}\right|^{2} (2 + 3 n_2) \gamma_2 \gamma_3 (\gamma_2 +
	n_2 \gamma_2 + \gamma_3) \gamma_4  +  \left| J_{1}\right|^{2} \big(n_2^2 \gamma_2^2 \gamma3
	\gamma_4 +  2 \left| J_{2}\right|^{2}  ((\gamma_2 + \gamma_3) \gamma_4 + n_2 \gamma_2 (\gamma_3 + 2
	\gamma_4)) \big)\Big)\bigg)\Bigg] = 0
	\label{eq:sol_luul}
\end{multline}
\end{widetext}
for the quantity $\left<\ket{l_{j}}\bra{u_{j}}\ket{u_{k}}\bra{l_{k}}\right>_{\text{st}}$
describing cooperative behavior between the structures. 
The stable solution is obtained from the part of Eq.~\eqref{eq:sol_luul} which is enclosed
by the rectangular brackets. 
To find the number of structures necessary for reaching the onset of cooperative behavior
$N_{\text{th}}\left( T_{1} \right)$
for a given temperature $T_{1}$ we solve
$\left<\ket{l_{j}}\bra{u_{j}}\ket{u_{k}}\bra{l_{k}}\right>_{\text{st}}=0$ to find
\begin{widetext}
\begin{multline}
	N_{\text{th}}\left( T_{1} \right)=\Bigg((\gamma_{1} + \gamma_{4}) \kappa \bigg(\left|
	J_{2}\right|^{2} (2 + 3 n_{2})
	\gamma_{1} \gamma_{2} \gamma_{3} (\gamma_{2} + n_{2} \gamma_{2} + \gamma_{3})
	\gamma_{4}\\
	+ \left| J_{1}\right|^{2} (n_{2}^2 \gamma_{1} \gamma_{2}^2 \gamma_{3} \gamma_{4}
	+ \left| J_{2}\right|^{2} (2 \gamma_{1} (\gamma_{2} + \gamma_{3}) \gamma_{4} +   n_{2} \gamma_{2} (\gamma_{1}
	\gamma_{3} +  4 \gamma_{1} \gamma_{4} + \gamma_{3} \gamma_{4})))\bigg)\Bigg)\\
	/\Bigg(g^2
	n_{2} \gamma_{2} \gamma_{3} (\gamma_{4} - \gamma_{1}) \left|J_{1}\right|^2
	\left|J_{2}\right|^2\Bigg)\ .
	\label{eq:N_th}
\end{multline}
\end{widetext}
The result is plotted in Fig.~\ref{fig:laser_photons}(b) (dashed line).
Similar calculations give the critical temperature $T_{1,\text{th}}$ for given $N$, see
Fig.~\ref{fig:laser_photons}(a).

\subsection*{Calculation of the spectrum}

Using the quantum regression theorem we obtain $\left< a^{\dagger}\left( t+\tau
\right)a\left( t \right) \right>$ which couples to $\left< \ket{u_{j}}\bra{l_{j}}\left(
t+\tau \right) a\left( t \right)\right>$. For $t\rightarrow \infty$ using stationarity we
can write
\begin{widetext}
\begin{align}
	\frac{\mathrm{d}}{\mathrm{d}\tau}\begin{pmatrix} \left< a^{\dagger}\left( \tau
		\right)a\left( 0 \right) \right>  \\ \left< \ket{u_{j}}\bra{l_{j}}\left( \tau
		\right) a\left( 0 \right)\right>  \end{pmatrix}=
		\begin{pmatrix} -\kappa & -\mathrm{i}gN\\
			-\mathrm{i} g \left( \left< \ket{u_{j}\bra{u_{j}}} \right>_{\text{st}}-  \left<
			\ket{l_{j}\bra{l_{j}}} \right>_{\text{st}}\right) & \mathrm{i}\Delta_{u}-\Gamma \end{pmatrix} \times \begin{pmatrix} \left< a^{\dagger}\left( \tau
		\right)a\left( 0 \right) \right>  \\ \left< \ket{u_{j}}\bra{l_{j}}\left( \tau
		\right) a\left( 0 \right)\right>  \end{pmatrix}
	\label{eq:correlation_equations}
\end{align}
\end{widetext}
with $\Gamma= \gamma_{3}n_{3}\left( T_{2} \right)+\gamma_{1}\left( 2 n_{1}\left( T_{2}
\right)+1 \right)+\gamma_{4}\left( n_{4}\left( T_{2} \right)+1 \right)$. The Laplace
transformation yields
\begin{align}
	\left< a^{\dagger}a \right>(s)=\frac{\left< a^{\dagger}a \right>_\text{st}\left(
	s-\mathrm{i}\Delta_{u}+\Gamma
	\right)-\mathrm{i}gN\left< \ket{u_{j}}\bra{l_{j}}a\right>_{\text{st}}}{\left( s+\kappa \right)\left(
	s-\mathrm{i}\Delta_{u}+\Gamma \right)-g^{2}N\left( \left< \ket{u_{j}}\bra{u_{j}}
	\right>_{\text{st}}- \left< \ket{l_{j}}\bra{l_{j}}
	\right>_{\text{st}}\right)}
	\label{eq:ada_s}
\end{align}
which allows to calculate the spectrum~\citep{meystre2007elements}.


\begin{thebibliography}{10}%
\makeatletter
\providecommand \@ifxundefined [1]{%
 \@ifx{#1\undefined}
}%
\providecommand \@ifnum [1]{%
 \ifnum #1\expandafter \@firstoftwo
 \else \expandafter \@secondoftwo
 \fi
}%
\providecommand \@ifx [1]{%
 \ifx #1\expandafter \@firstoftwo
 \else \expandafter \@secondoftwo
 \fi
}%
\providecommand \natexlab [1]{#1}%
\providecommand \enquote  [1]{``#1''}%
\providecommand \bibnamefont  [1]{#1}%
\providecommand \bibfnamefont [1]{#1}%
\providecommand \citenamefont [1]{#1}%
\providecommand \href@noop [0]{\@secondoftwo}%
\providecommand \href [0]{\begingroup \@sanitize@url \@href}%
\providecommand \@href[1]{\@@startlink{#1}\@@href}%
\providecommand \@@href[1]{\endgroup#1\@@endlink}%
\providecommand \@sanitize@url [0]{\catcode `\\12\catcode `\$12\catcode
  `\&12\catcode `\#12\catcode `\^12\catcode `\_12\catcode `\%12\relax}%
\providecommand \@@startlink[1]{}%
\providecommand \@@endlink[0]{}%
\providecommand \url  [0]{\begingroup\@sanitize@url \@url }%
\providecommand \@url [1]{\endgroup\@href {#1}{\urlprefix }}%
\providecommand \urlprefix  [0]{URL }%
\providecommand \Eprint [0]{\href }%
\providecommand \doibase [0]{http://dx.doi.org/}%
\providecommand \selectlanguage [0]{\@gobble}%
\providecommand \bibinfo  [0]{\@secondoftwo}%
\providecommand \bibfield  [0]{\@secondoftwo}%
\providecommand \translation [1]{[#1]}%
\providecommand \BibitemOpen [0]{}%
\providecommand \bibitemStop [0]{}%
\providecommand \bibitemNoStop [0]{.\EOS\space}%
\providecommand \EOS [0]{\spacefactor3000\relax}%
\providecommand \BibitemShut  [1]{\csname bibitem#1\endcsname}%
\let\auto@bib@innerbib\@empty
\bibitem [{\citenamefont {Lamb~Jr}(1964)}]{lamb1964theory}%
  \BibitemOpen
  \bibfield  {author} {\bibinfo {author} {\bibfnamefont {W.}~\bibnamefont
  {Lamb~Jr}},\ }\href@noop {} {\bibfield  {journal} {\bibinfo  {journal} {Phys.
  Rev}\ }\textbf {\bibinfo {volume} {134}},\ \bibinfo {pages} {A1429} (\bibinfo
  {year} {1964})}\BibitemShut {NoStop}%
\bibitem [{\citenamefont {Boukobza}\ and\ \citenamefont
  {Tannor}(2006)}]{boukobza2006thermodynamic}%
  \BibitemOpen
  \bibfield  {author} {\bibinfo {author} {\bibfnamefont {E.}~\bibnamefont
  {Boukobza}}\ and\ \bibinfo {author} {\bibfnamefont {D.}~\bibnamefont
  {Tannor}},\ }\href@noop {} {\bibfield  {journal} {\bibinfo  {journal}
  {Physical Review A}\ }\textbf {\bibinfo {volume} {74}},\ \bibinfo {pages}
  {063822} (\bibinfo {year} {2006})}\BibitemShut {NoStop}%
\bibitem [{\citenamefont {Youssef}\ \emph {et~al.}(2010)\citenamefont
  {Youssef}, \citenamefont {Mahler},\ and\ \citenamefont
  {Obada}}]{youssef2010quantum}%
  \BibitemOpen
  \bibfield  {author} {\bibinfo {author} {\bibfnamefont {M.}~\bibnamefont
  {Youssef}}, \bibinfo {author} {\bibfnamefont {G.}~\bibnamefont {Mahler}}, \
  and\ \bibinfo {author} {\bibfnamefont {A.}~\bibnamefont {Obada}},\
  }\href@noop {} {\bibfield  {journal} {\bibinfo  {journal} {Physica E:
  Low-dimensional Systems and Nanostructures}\ }\textbf {\bibinfo {volume}
  {42}},\ \bibinfo {pages} {454} (\bibinfo {year} {2010})}\BibitemShut
  {NoStop}%
\bibitem [{\citenamefont {Graham}\ and\ \citenamefont
  {Haken}(1970)}]{graham1970laserlight}%
  \BibitemOpen
  \bibfield  {author} {\bibinfo {author} {\bibfnamefont {R.}~\bibnamefont
  {Graham}}\ and\ \bibinfo {author} {\bibfnamefont {H.}~\bibnamefont {Haken}},\
  }\href@noop {} {\bibfield  {journal} {\bibinfo  {journal} {Zeitschrift
  f{\"u}r Physik A Hadrons and Nuclei}\ }\textbf {\bibinfo {volume} {237}},\
  \bibinfo {pages} {31} (\bibinfo {year} {1970})}\BibitemShut {NoStop}%
\bibitem [{\citenamefont {Kumar}\ \emph {et~al.}(2010)\citenamefont {Kumar},
  \citenamefont {Chan}, \citenamefont {Hu},\ and\ \citenamefont
  {Reno}}]{kumar20101}%
  \BibitemOpen
  \bibfield  {author} {\bibinfo {author} {\bibfnamefont {S.}~\bibnamefont
  {Kumar}}, \bibinfo {author} {\bibfnamefont {C.}~\bibnamefont {Chan}},
  \bibinfo {author} {\bibfnamefont {Q.}~\bibnamefont {Hu}}, \ and\ \bibinfo
  {author} {\bibfnamefont {J.}~\bibnamefont {Reno}},\ }\href@noop {} {\bibfield
   {journal} {\bibinfo  {journal} {Nature Physics}\ }\textbf {\bibinfo {volume}
  {7}},\ \bibinfo {pages} {166} (\bibinfo {year} {2010})}\BibitemShut {NoStop}%
\bibitem [{\citenamefont {Epstein}\ \emph {et~al.}(1995)\citenamefont
  {Epstein}, \citenamefont {Buchwald}, \citenamefont {Edwards}, \citenamefont
  {Gosnell},\ and\ \citenamefont {Mungan}}]{epstein1995observation}%
  \BibitemOpen
  \bibfield  {author} {\bibinfo {author} {\bibfnamefont {R.}~\bibnamefont
  {Epstein}}, \bibinfo {author} {\bibfnamefont {M.}~\bibnamefont {Buchwald}},
  \bibinfo {author} {\bibfnamefont {B.}~\bibnamefont {Edwards}}, \bibinfo
  {author} {\bibfnamefont {T.}~\bibnamefont {Gosnell}}, \ and\ \bibinfo
  {author} {\bibfnamefont {C.}~\bibnamefont {Mungan}},\ }\href@noop {}
  {\bibfield  {journal} {\bibinfo  {journal} {Nature}\ }\textbf {\bibinfo
  {volume} {377}},\ \bibinfo {pages} {500} (\bibinfo {year}
  {1995})}\BibitemShut {NoStop}%
\bibitem [{\citenamefont {Carmichael}(1993)}]{carmichael1993open}%
  \BibitemOpen
  \bibfield  {author} {\bibinfo {author} {\bibfnamefont {H.}~\bibnamefont
  {Carmichael}},\ }\href@noop {} {\emph {\bibinfo {title} {An open systems
  approach to quantum optics: lectures presented at the Universit{\'e} libre de
  Bruxelles, October 28 to November 4, 1991}}},\ Vol.~\bibinfo {volume} {18}\
  (\bibinfo  {publisher} {Springer},\ \bibinfo {year} {1993})\BibitemShut
  {NoStop}%
\bibitem [{\citenamefont {Kubo}(1962)}]{kubo1962generalized}%
  \BibitemOpen
  \bibfield  {author} {\bibinfo {author} {\bibfnamefont {R.}~\bibnamefont
  {Kubo}},\ }\href@noop {} {\bibfield  {journal} {\bibinfo  {journal} {Journal
  of the Physical Society of Japan}\ }\textbf {\bibinfo {volume} {17}},\
  \bibinfo {pages} {1100} (\bibinfo {year} {1962})}\BibitemShut {NoStop}%
\bibitem [{\citenamefont {Meiser}\ \emph {et~al.}(2009)\citenamefont {Meiser},
  \citenamefont {Ye}, \citenamefont {Carlson},\ and\ \citenamefont
  {Holland}}]{meiser2009prospects}%
  \BibitemOpen
  \bibfield  {author} {\bibinfo {author} {\bibfnamefont {D.}~\bibnamefont
  {Meiser}}, \bibinfo {author} {\bibfnamefont {J.}~\bibnamefont {Ye}}, \bibinfo
  {author} {\bibfnamefont {D.}~\bibnamefont {Carlson}}, \ and\ \bibinfo
  {author} {\bibfnamefont {M.}~\bibnamefont {Holland}},\ }\href@noop {}
  {\bibfield  {journal} {\bibinfo  {journal} {Physical review letters}\
  }\textbf {\bibinfo {volume} {102}},\ \bibinfo {pages} {163601} (\bibinfo
  {year} {2009})}\BibitemShut {NoStop}%
\bibitem [{\citenamefont {Meystre}\ and\ \citenamefont
  {Sargent}(2007)}]{meystre2007elements}%
  \BibitemOpen
  \bibfield  {author} {\bibinfo {author} {\bibfnamefont {P.}~\bibnamefont
  {Meystre}}\ and\ \bibinfo {author} {\bibfnamefont {M.}~\bibnamefont
  {Sargent}},\ }\href@noop {} {\emph {\bibinfo {title} {Elements of quantum
  optics}}}\ (\bibinfo  {publisher} {Springer Verlag},\ \bibinfo {year}
  {2007})\BibitemShut {NoStop}%
\end{thebibliography}
\end{document}